\documentclass[aps,prd,onecolumn,groupedaddress,nofootinbib,amssymb]{revtex4}
\usepackage[T1]{fontenc}
\usepackage[latin1]{inputenc}
\usepackage{graphicx}
\usepackage[english]{babel}
\usepackage{graphicx}
\usepackage{bm}
\usepackage{amsmath}
\usepackage{amssymb}
\usepackage{amsfonts}
\usepackage{epsfig}
\usepackage{colordvi}
\usepackage{color}

\begin{document}
\title{Gamma Ray Bursts Scaling Relations  to test cosmological models}
\author{S.  Capozziello$^{1,2}$,  L.  Consiglio$^{1,2}$,  M.  De Laurentis$^{1,2}$, 
  G.  De Rosa$^{1,2}$, C. Perna$^3$,   D. Vivolo$^{1,2}$ }
\affiliation{\it $^1$Dipartimento di Scienze Fisiche, Universit\'a
di Napoli {}``Federico II'', and
$^2$INFN Sez.  di Napoli, \\ Compl. Univ. di
Monte S. Angelo, Edificio G, Via Cinthia, I-80126, Napoli, Italy,\\
$^3$INAF - Osservatorio Astronomico di Monte Mario, Viale del Parco Mellini 84, I-00136 Roma, Italy.} 
% $^4$INAF - Istituto di Radioastronomia, Via P. Gobetti, 10, I-40129 Bologna, Italy.} 
\date{\today}

\begin{abstract}
Gamma ray burst (GRBs) can be used to constrain cosmological
parameters from medium up to very high redshift. These powerful
systems could be the further reliable distance indicators  after SNeIa
supernovae. We consider GRBs samples to achieve the luminosity
distance to redshift relation and derive the values of the
cosmographic parameters considering several possible scaling
relations.  GRBs, if calibrated by SNeIa, seem reliable as
distance indicators and give cosmographic parameters in good agreement
with the $\Lambda$CDM model. GRBs correlations with neutrino and
gravitational wave signals are also investigated in view of high energy neutrino experiments and gravitational wave detectors as LIGO-VIRGO.
 A discussion on the
GRB afterglow curve up to the visible and radio wavelengths is
developed considering the possibility to use the Square Kilometer
Array (SKA) telescope to achieve the first GRB-radio survey.  
\end{abstract}

\maketitle

\noindent \textbf{PACS} 05.45-a, 52.35.Mw, 96.50.Fm.

\noindent \textbf{Keywords:} Gamma Ray Bursts;  Cosmology; Gravitational Waves; Neutrinos.

%\label{lastpage-01}

%%%%%%%%%%%%%%%%
\section{Introduction}
%%%%%%%%%%%%%%%%%%%%%%%%%

Observational data collected in the last fifteen
years, as the anisotropy and polarization spectra of the cosmic
microwave background radiation (CMBR) \cite{Boom00,QUAD09,WMAP7}, the
large scale structure of galaxy redshift surveys
\cite{D02,P02,Sz03,H03}, the measurements of Baryonic Acoustic Oscillations (BAO,
\cite{Eis05,P10}) and the Hubble diagram derived from Supernovae Type Ia (SNeIa)
\cite{Union,H09,SNeIaSDSS}), strongly support the cosmological
picture   of a spatially flat universe with a subcritical matter content
$(\Omega_M \sim 0.3)$   undergoing an accelerated phase of
expansion.
 While the observational overview is now firmly established,
the search for the motivating theory is, on the contrary, still dawning 
despite of several efforts and the abundance of models
proposed during these years. The question is not the
lack of a well established theory, but the presence of too many viable
candidates, ranging from the classical cosmological constant
\cite{CPT92,SS00}, to scalar fields \cite{PR03,Copeland06} and higher
order gravity theories \cite{CF08,FS08,ND08,dFT10,report,francaviglia}, all of them being
more or less capable of fitting the available data.

As usual, adding further reliable data is the best strategy to put
order in the confusing abundance of theoretical models. In particular,
the extension of the observed Hubble Diagram (HD) to higher redshift $z$, would allow to
trace the universe background evolution up to the transition regime
from the accelerated dark energy era to the decelerated matter
dominated phase. Moreover, being the distance modulus
 related to the luminosity distance and depending  on the dark energy equation of state, one should go to large $z$ in order to discriminate among
different models when these predict similar
curves at lower redshift. Unfortunately, SNeIa are not well suitable
for this aim since their current Hubble diagram go back to
$z_{max} \sim 1.4\div1.7$ and does not extend further than $z \simeq 2$ even
for excellent space based experiments such as SNAP \cite{SNAP}.
Unlike GRBs, due to their enormous, almost instantaneous energy release,  stand out as ideal candidates to explore further
redshift, the farthest one today detected at $z = 8.3$
\cite{Salvaterra2009}. The wide range spanned by their peak energy
makes them everything but standard candles; anyway the existence of many
observationally motivated correlations between redshift dependent
quantities and rest frame properties
\cite{Amati08,FRR00,G04,liza05} offers the intriguing possibility
of turning GRBs into standardizable objects as SNeIa. 

Many attempts to use GRBs as cosmological distance indicators tools
have been already performed (see, {\it e.g.}, \cite{fir06,L08,LWZ09,QiLu09,Izzo08,Izzo09}  and refs. therein)
showing the potentiality of GRBs as cosmological probes.

It is mandatory to remind that the possibility offered by GRBs to track
the HD deep into the matter dominated epoch does not come for
free. Two main problems are actually still to be fully
addressed. First, missing a local GRBs sample, all the possible
correlations have to be calibrated assuming a fiducial cosmological
model to estimate the redshift dependent quantities. As a consequence,
the so called {\it circularity problem} comes out, that is to say one wants
to use GRBs scaling relations to constrain the basic cosmology,
but needs the basic cosmology to get the scaling
relations \cite{perillo}.
A well behaved distance indicator should be not only visible to high
$z$ and characterized by scaling relations with as small as possible intrinsic
scatter, but its physics should be well understood from a theoretical
point of view. Presently, there is no  full 
understanding of the GRBs scaling relations so that, as a dangerous
drawback, one cannot anticipate whether the calibration parameters
are redshift dependent or not. Since we cannot refer to specific theoretical models, one tries to address this problem in a phenomenological way.

This review, without claim of completeness, is an attempt in this direction. We will try to summarize some scaling relations and interesting features (most of them already present in literature) that could result useful to standardize GRBs in view of cosmology.  The paper is organized as follows. The main features of GRB phenomena are  sketched in Sec.2.  The so called {\it fireball model} is shortly reviewed in Sec.3.  Cosmology with GRBs is widely discussed in Sec.4. Here we recall the main scaling relations that could be useful in order to figure out GRBs as possible standard cosmological indicators. In particular, we discuss the correlation analysis in view of testing cosmological models. Implication for particle astrophysics (i.e. high energy neutrinos) and gravitational radiation are taken into account in Sec.5. Sec.6 is devoted to the GRB radio emission which could be extremely important to accomplish the luminosity curve of these objects in the whole electromagnetic spectrum.  Conclusions are drawn in Sec.7.

%%%%%%%%%%%%%%%%%%%%%%%%%%%%%%%%%%%%%%%%%%%%%%%%
\section{The Gamma-Ray Burst phenomenon}
%%%%%%%%%%%%%%%%%%%%%%%%%%%%%%%%%%%%%%%%

GRBs are extremely powerful flashes of $\gamma$-rays which occur approximately
once per day and are isotropically distributed over the
sky \cite{Meszaros:2006rc}. The variability of the bursts on time
scales as short as a millisecond and indicates that the sources are very
compact, while the identification of host galaxies and the measurement
of redshifts for more than $100$ bursts have shown that GRBs are of
extra-galactic origin. GRBs are grouped into two broad classes by
their characteristic duration and spectral hardness: {\it long} and {\it short} GRBs \cite{ck93,Gehrels2006}. 
Moreover,  cosmological GRBs are very likely powerful sources of high
energy neutrinos and gravitational waves.
According to the current interpretation of GRB phenomenology, the $\gamma$-ray emission is due to the dissipation of the kinetic energy of a relativistically expanding fireball \cite{Meszaros:2006rc}. 
The physical conditions allow protons to be accelerated to energies  greater than
$10^{20}$ eV according to the Fermi mechanism \cite{Waxman1,Vietri,Waxman2,Bahcall}.
Such protons  cannot avoid interactions with fireball
photons, starting a photo-meson production process that generate very high energy neutrinos and $\gamma$-rays \cite{Waxman3}.

The occurrence of gravitational wave bursts (GWBs) associated with GRBs
is a natural consequence of current models for the central engine \cite{3}. For instance, GRBs can be produced by  classes of supernovae, known as {\it collapsars} or 
{\it hypernovae}, when a massive star collapses to form a spinning black hole; in the meantime the remaining core materials
form an accreting torus with high angular momentum \cite{4, 5}. 
Another interesting scenario is
a neutron star and black hole coalescing system ($\sim 7M_\odot$) where the disruption of the neutron star, caused by the rapidly rotating black hole, will also form a torus emitting a large amount of energy ($\sim 0.1M_{\odot}c^2$) both in gravitational and electromagnetic waves \cite{4}.

According to the fireball model, GRB afterglows are the result of a shock pushed into the ambient medium by an extremely relativistic outflow from the GRB. A conducting fireball expanding at relativistic speed into an ambient magnetic field generates a rapidly changing electric current which emits coherent electromagnetic radiation at radio frequencies. The critical frequency (upper limit of the emission) strongly depends on the Lorentz factor of the expansion.
Wide radio observations of GRB afterglows would provide essential and unique information to constrain the physical models by completing the coverage of the spectral energy distribution and following the behavior up to much later times than any other wavelength.

Due to the above processes, GRBs are the most powerful phenomena of the universe located at cosmological distances \cite{Meszaros:2006rc,piran04}. They were discovered in the late $60$'s by the military VELA satellites, as  $X$-ray bursts and $\gamma$-ray photons occurring at random place and date in the sky \cite{kle73}. Further observations led to the discovery of slowly fading $X$-ray, optical and radio afterglows following the burst of photons of the {\em prompt} phase, and the identification of host galaxies at cosmological distances. 

 Specifically, the prompt emission phase of long bursts lasts typically $20$ seconds, and more than two seconds. The short bursts have a shorter duration, typically of $0.2$ seconds \cite{dez92,kou93}. Both classes present an afterglow emission, rather similar \cite{short}. The difference between the prompt phases of short and long GRBs derives from  the nature of the progenitor.
The progenitors of most short GRBs  are widely thought to be mergers of neutron star
 binaries (NS-NS) or neutron star-black hole (NS-BH) binaries \cite{nakar-2007}.
 A small fraction (up to $\simeq$15\%) of short GRBs
are also thought to be due to giant flares from a local distribution
of soft-gamma repeaters \cite{duncan92,NaGaFo:06,Chapman:2007xs}. Long GRBs, on the other hand, are associated with core-collapse
Supernovae \cite{galama98,hjorth03,Ma_etal:04,Campana:2006qe}. As said above, both the merger and supernova scenarios
result in the formation of a stellar-mass black hole with accretion
disk. The emission of
gravitational radiation and neutrinos are expected in this process \cite{fryer99,Cannizzo:2009qv}. 
GRBs are also classified on the basis of other electromagnetic properties. One can,  for example, consider the optical brightness of the afterglow, distinguishing {\em dark} bursts, having no optical afterglow emission \cite{dep03,horst}; or {\em $X$-ray Flashes} (which have no emission in the $\gamma$-rays, the prompt being reduced to a burst of soft $X$-ray photons) \cite{hei00}. These classifications are  based on observations and can be related to the properties of the medium in which the burst develops or to the geometry of the explosion \cite{dep03,gen07,gendre}.

%%%%%%%%%%%%%%%%%
\section{The fireball model}
%%%%%%%%%%%%%%%%

The {\it fireball model} is based on an "central engine" highly energetic
which produces a relativistic particle outflow. Since the opacity is
very high, this "engine" is well hidden from direct observations and it
is very difficult to determine what is the source of this mechanism
from present observations. Also the afterglow discovery does not
add further information in this direction, leading only to some circumstantial
evidences on the nature of the sources.  The energy from the central
source is transported relativistically to distances larger than
$10^{16}$cm where the system is optically thin. The fireball particles
flux is not emitted at a constant rate by the central engine but two
jets back to back occur  in the near of the central engine.  Such jets are made up
of photons, baryons and $e^-/e^+$ pairs that accelerate the surrounding
material by forming concentric and consecutive shells. During the
acceleration process, some shells are emitted with a higher velocity
respect to others.  Faster shells interact with slower ones causing an
inelastic shock which accelerates electrons producing a gamma emission
at high energy.  This mechanism called {\em internal} shock powers the
GRB itself producing the prompt emission \cite{ree92}.  Subsequently,
the fireball, during its expansion, bumps into external medium which
can be the interstellar medium or the dense stellar wind produced by
the progenitor of GRB.  External shocks arise due to the interaction
of the relativistic matter with the surrounding matter, and cause the
GRB afterglow. These shocks or blast waves are the relativistic
analogues of supernova remnants.  The shock will produce a magnetic
field within the top of the jet, so that the electrons will start to
emit synchrotron radiation.  When the external forward shock is
formed, a reverse shock is produced moving back into the ejecta. This
reverse shock can produce a bright optical flash about one minute
after the burst, and a radio flare about one day after the burst. The
brightness of the reverse shock emission decays very rapidly, after
which the forward shock dominates. The relativistic outflow is not
spherical but collimated. Since the shock decelerates while it is
sweeping up mass, a few days after the burst, the synchrotron emission
angle of the electrons becomes equal to the collimation angle of the
outflow, the so-called jet-break time. After this moment the
collimated outflow spreads and becomes spherical after a few months
after the burst. At the same time that the outflow becomes
spherical, the blast wave becomes sub-relativistic and will eventually
enter the classical regime. GRBs occur at a rate of about one per
$10^6$ years per galaxy \cite{56} and the total energy is $\sim10^{52}$
ergs. These estimates are obtained assuming isotropic emission.  The
jet beaming angle $\theta_{jet}$ can changes the rate estimation by a
factor $\displaystyle{\frac{4\pi}{\theta^2_{jet}}}$ in the rate, and
by a factor of $\displaystyle{\frac{\theta^2_{jet}}{4\pi}}$ in the
fireball total energy.  The luminosity per unit solid angle along the
jet axis is different from the one emitted on axis. A strongly
off-axis position of a detector would not allow the prompt GRBs
observation so that it would be possible to see (in case of a long
GRB) only the late afterglow \cite{dal05}.  A further constraint on
fireball model is the inner engine capability of accelerating
$\sim10^{-5}M_{\odot}$ to relativistic energies.

One can imagine
various scenarios in which $10^{52}$ ergs are generated within a short
time. The condition that such energy should be converted into a
relativistic flow is quite difficult since it needs a system with a
very low baryonic load. This requirement prefers models based on
electromagnetic energy transfer or electromagnetic energy generation
as these could more naturally satisfy this condition (see \cite{267}).

%%%%%%%%%%%%%%%%%%%%%%%%%%%%%%%%%%%%%%%%%%%
\section{Cosmology by GRBs}
%%%%%%%%%%%%%%%%%%%%%%%%%%%%%%%%%%%%%%%%%%%%

 The cosmological origin of GRBs has been confirmed by
several spectroscopic measurements of their redshifts, distributed in the range
$z \in  0.1\div 8.3$. This  property makes GRBs suitable for investigating
the far universe. Indeed, they can be used to constrain the geometry of the
present day universe and the nature and evolution of dark energy by testing the
cosmological models in a redshift range hardly achievable by other cosmological
probes. Moreover, the use of GRBs as cosmological tools could probe the
ionization history of the universe, the intergalactic medium properties and
the formation of massive stars in the early universe. 
 The fact that GRBs are detected up to very high redshifts makes it possible an attempt 
to  use them as standard candles in
a similar way as SNeIa, because they have
a very wide range of isotropic equivalent luminosities
and energy outputs. Several suggestions have been made
to calibrate them as better standard candles by using
correlations between various properties of the prompt
emission, and in some cases also the afterglow emission.
While there is good motivation for such cosmological applications of GRBs, there are many practical difficulties. 
 
 Indeed, a serious problem that arises is the intrinsic scarcity of
 the nearby events which introduces a bias towards low/high values of
 GRB observables. As a consequence, it is not immediate to extrapolate
 the correlations to low-$z$ events.  A further problem is related to
 the fact that, due to the unknown flux limit, the GRB ensemble suffer
 from the well known degradation of sampling as a function of
 redshift.  One might also expect a significant evolution of the
 intrinsic properties of GRBs with redshift (also between intermediate
 and high redshifts) which can be hard to decouple from cosmological
 effects.  Finally, in order to calibrate  the observed
 correlations, it is mandatory to assume a cosmological model
 (luminosity distance vs redshift) in order to convert the observed
 bolometric peak flux $P_{bolo}$ or bolometric fluence to isotropic
 absolute luminosity $L$ or to a total collimation corrected energy
 $E_\gamma$. The use of a cosmological model to perform the
 calibration, in turn, creates a circularity problem and a model
 dependence of the obtained calibration.
  
Despite of these difficulties, the potential advantages to obtain approximate standard candles at high redshifts have generated an intense 
activity in the direction to test GRBs as
cosmological indicators \cite{fir06,liza05} as well as  to constrain
cosmological parameters by them \cite{S03,schaefer,zha04,Amati08}. 
This means that GRB cannot be alternative to the SNeIa or to other cosmological probes, but they can be complementary to them due to their wide redshift distribution and  evolution properties. The goal is to use them to remove  the   degeneracies in the values of cosmological parameters, today obtained only at medium and low redshift (see for example \cite{wangd,wangd1}).  
 
 %%%%%%%%%%%%%%%%%%%%%%
 \subsection{The standard candles test}
%%%%%%%%%%%%%%%%%%%%%%
The standard methods 
to test  cosmological models,  are: 
\begin{itemize}
\item the luminosity distance test (mainly applied to SNeIa); 
\item the angular size distance test (used also for  Cosmic Microwave Background (CMB) anisotropies);
\item the volume test based on galaxy number count.
\end{itemize}
In general, given an object with a certain luminosity $L$, not evolving with
cosmic time, and measured flux $F$, one can define its luminosity
distance $\displaystyle{D_L = \sqrt{\frac{L}{4\pi F(z) }}}$ which is
related to the radial coordinate $r$ of the Friedman-Robertson-Walker
metric by $\displaystyle{D_L = \frac{r}{a(\tau)} = r (1 + z)}$, where
$a(\tau)$ is the scale factor as a function of the comoving time,
$\tau = t H_0$. $D_L$ depends on the expansion history and curvature
of the universe through the radial coordinate $r$. By measuring the
flux of {\it 'standard candles'} as a function of redshift, $F(z)$,
the luminosity distance test is achieved by comparing $D_L$, obtained from
the flux measurement, with $D_L({\bar p})$ predicted by the
cosmological model, where ${\bar p}$ is a set of cosmological
parameters ({\it e.g.}, the matter density parameter $\Omega_M$,  the cosmological density parameter $\Omega_\Lambda$ and
the normalized Hubble constant $h$ \cite{gravitation,ghi}).
 This test has been largely used with SNeIa  \cite{ghi}. The high peak luminosity ({\it i.e.} $\sim 10^{10}M_{\odot})$ of SNeIa makes them detectable above $z >1$. 

Several intrinsic correlations between temporal or spectral properties and GRB isotropic energetics and luminosities have been considered as a possibility to use GRBs as cosmological indicators \cite{basilakos}. The constraints on the cosmological parameters obtained by the luminosity distance test applied to GRBs are less severe than those obtained by SNeIa. This is mainly due to the presently still limited number of GRBs with well determined prompt and afterglow properties that can be used as standard indicators.

 %%%%%%%%%%%%%%%%%%%%%%%%%%%%%%%%%%%%%%%
 \subsection{GRB luminosities and energetics}
%%%%%%%%%%%%%%%%%%%%%%%%%%%%%%%%%%%%%%
 
 The luminosity $L$ and the energy $E$ of GRBs with well know
 redshifts can be evaluated by the peak flux $P$ and the flux
 integrated over the burst duration ({\it i.e.} the fluence $S$).  If
 GRBs emit isotropically, the energy radiated during their prompt
 phase is $\displaystyle{ E_{iso} =\frac{4\pi D_{L}^{2}S}{(1+z)}}$
 where the term $(1 + z)$ accounts for the cosmological time dilation
 effect while the isotropic luminosity is $L_{iso} = 4\pi D_{L}^{2}
 P$.  The bolometric luminosity is often computed by combining the
 peak flux (relative to the peak of the GRB light curve) with the
 spectral data derived from the analysis of time integrated
 spectrum. As discussed in \cite{56}, this is strictly correct only if
 the GRB spectrum does not evolve in time during the burst \cite{57}.  
 A considerable reduction of
  dispersion of GRBs energetics has been found in \cite{58} (later
 confirmed in \cite{59}) when they are corrected for the collimated
 geometry of these sources.  Theoretical considerations on the extreme
 GRB energetics under the hypothesis of isotropic emission
 \cite{60,61} led to the hypothesis  that, similarly to other sources, also
 GRBs might be characterized by a jet.  In the classical scenario, the
 presence of a jet \cite{62,63,64} affects the afterglow light curve
 which should present an achromatic break a few days after the burst.
 The observation of the afterglow light curves allows  to
 estimate the jet opening angle $\theta_{jet}$ from which the
 collimation factor can be computed, {\it i.e.} $f = (1- \cos
 \theta_{jet} )$. This geometric correction factor, applied to the
 isotropic energies \cite{58}, led to a considerable reduction of the
 GRB energetics and of their dispersion.
 
%%%%%%%%%%%%%%%%%%%%%%%%%%%%%%%%%%%%%%% 
 \subsection{The intrinsic correlation of GRBs}
 %%%%%%%%%%%%%%%%%%%%%%%%%%%%%%%%%%%
 
 Several empirical relations between observable properties of the
 prompt emission have been discovered during the recent years.  The
 luminosity relations are connections between measurable parameters of
 the light curves and/or spectra with the GRB luminosity.  The
 calibration consists in a fit of the luminosity indicator versus the
 luminosity in logarithmic scale.  This calibration process needs to
 know  burst's luminosity distance in order to convert $P_{bolo}$
 to $L$ and this can be done only for bursts with measured redshifts.
 The crucial point is that  conversion from  observed redshift
 to luminosity distance requires some adopted cosmological
 parameters to be calibrated separately.
 
If we are interested in calibration for purposes of GRB physics, then it will be fine to adopt the calibration from some fiducial cosmology.  But if we are interested in testing the cosmology, then we have to use the calibration for each individual cosmology being tested.  For a particular cosmology, the theoretical shape of HD has to be compared with the observed shape when the burst distances are calculated based on calibrations for that particular cosmology. Thus, any test of cosmological models with a GRB-HD will be a simultaneous fit of the parameters in the calibration curves and the cosmology.
In the following subsections these correlations are summarized.
 
 %%%%%%%%%%%%%%%%%%%%%%%%%%%%%%
  \subsubsection{The lag-luminosity correlation $\tau-L_{iso}$}
  %%%%%%%%%%%%%%%%%%%%%%%%%%%%%%%%%%%

Results from BATSE satellite \cite{batse} on the analysis of the light curves of GRBs led to the
discovery of spectral lags, {\it i.e.} the difference in 
arrival time of high and low energy photons. It is assumed positive when
high energy photons arrive earlier than the low energy ones.  Usually
the spectral lag is extracted between two energy bands in the observed
reference frame. Since GRBs are redshift dependent, the two energy
bands extracted can refer to a different couple of energy bands in
the GRBs absolute frame; in this way energy depends on the spectral
lag considered.  Time lags typically range between $0.01$ and
$0.5$s (even few second lags have been observed \cite{66}) and there
is no evidence of any trend, within multi peaked GRBs, between the lags
of the initial and the latest peaks \cite{67}. It has been proposed
that  lags are a consequence of the spectral evolution \cite{68},
typically observed in GRBs \cite{69}, and they have been interpreted
as due to radiative cooling effects \cite{70}. Other interpretations
concern geometric and hydrodynamic effects \cite{71,72} within the
standard GRB model.  In particular, the analysis of GRB temporal
properties  with known redshifts revealed a tight correlation
between their spectral lags $(\tau)$ and the luminosity $(L_{iso})$
\cite{73}.  Furthermore, the $\tau-L_{iso}$ correlation has been used
as a pseudo-redshift indicator to estimate $z$ for a large population
of GRBs \cite{80} and also to study the GRB population properties,
like as jet opening angle, luminosity function and redshift
distribution within a unifying picture \cite{66}.
 
 %%%%%%%%%%%%%%%%%%%%%%%%%%%%%%%%%%%%
 \subsubsection{The variability-luminosity correlation $V-L_{iso}$}
%%%%%%%%%%%%%%%%%%%%%%%%%%%%%%%%%%%%%%
GRB temporal structure  shows several shapes: they exhibit a variation from a single smooth pulse to very high complex light curves with a lot of fancy pulses with different amplitudes time duration.  
 Also the afterglow emission shows some variability on timescales \cite{83,84}. The analysis of large samples of bursts also showed the existence of a correlation between the GRB observer frame intensity and its variability \cite{81}. Several scenarios have been proposed to provide an explanation of GRBs temporal variability. The most accredited mechanism asserts that light curves variability would be due to the irregularity of the central engine. Alternatively, an external origin of the observed variability as due to the shock formation by the interaction of the relativistically expanding fireball and variable size interstellar medium clouds  is proposed in \cite{87}.
Fenimore and Ramirez-Ruiz \cite{88} and Reichart et al., \cite{89}  found a correlation between GRB luminosities $L_{iso}$ and their variability $V$: more luminous bursts have a more variable light curve. The $V-L_{iso}$ correlation has been recently updated \cite{90} with a sample of $31$ GRBs with measured redshifts. This correlation has also been tested \cite{91} with a large sample of $155$ GRBs with only a pseudo-redshift estimate (from the lag-luminosity correlation \cite{80,perillo}). An even tighter correlation ({\it i.e.} with a reduction of a factor $3$ of its scatter) has been derived \cite{92} by slightly modifying the definition of the variability first proposed in \cite{89}. Recently Zhang and Yan \cite{LZ} proposed Internal- Collision-Induced-Magnetic Reconnection and Turbulence  model  of GRBs prompt emission in the Poynting-flux-dominated regime. This model is based on a central engine with powered, magnetically dominated outflow which self interacts and triggers fast magnetic turbulent reconnection that powers the observed GRBs. This model has two variability components: one slow component is related to the central engine activity which is responsible of the visually apparent broad pulses in GRBs light curves;  the other fast one is associated with  relativistic magnetic turbulence responsible of the faster variabilities overlapping the broad pulses.
It is fundamental to use rigorous approaches to study GRBs light curves in order to investigate  superpositions of multiple variability components. Power density spectrum  is, however,  the most used tool to study the temporal behavior of varying astronomical objects \cite{Gao}.

%%%%%%%%%%%%%%%%%%%%%%%%%%%%%%%%%%%%%%%%%%%%%%%%%%  
 \subsubsection{The spectral peak energy-isotropic energy correlation $E_{peak}-E_{iso}$}
%%%%%%%%%%%%%%%%%%%%%%%%%%%%%%%%%%%%%%%%%%%%%%%%%%%%

The  $E_{peak}-E_{iso}$ relation is  one of the most latest intriguing discovery related to  GRBs and could have several implications on the  understanding  of the emission mechanism of the burst and on the possibility to use GRBs for cosmology.
Even before  large samples of burst with measured redshift were available, it was suggested that the $E_{peak}-E_{iso}$ were correlated \cite{95}. Hereafter the measurements of some redshift Amati et al. \cite{96} reported the correlation  between the isotropic equivalent bolometric energy output in $\gamma$-rays, $E_{iso}$, and the intrinsic peak energy of the $\nu F_\nu$ spectrum, $E_{peak}$. This result was based on a sample of $12$ BeppoSAX bursts \cite{beppo} with known redshifts. Ten additional bursts detected by HETE II (\cite{34,53,97}) supported this result and extended it down to $E_{iso}\sim 10^{49}$ ergs (see also \cite{ghi}).
 The $E_{peak}-E_{iso}$ correlation has been updated with a sample of $43$ GRBs (comprising also $2$XRF) by  estimating $z$ and the spectral properties \cite{55}. 
 The theoretical interpretations of  $E_{peak}-E_{iso}$ correlation attribute it to geometrical effects due to the jet viewing angle with respect to a ring-shaped emission region \cite{98,99} or with respect to a multiple sub-jet model structure which also takes into account for the extension of the above correlation to the $X$-ray rich (XRR) and XRF classes \cite{100,101}. A different  explanation for the $E_{peak}-E_{iso}$ correlation concerns the dissipative mechanism responsible for the prompt emission \cite{102}: if the peak spectral energy is interpreted as the fireball photospheric thermal emission comptonized by some mechanism ({\it e.g.} magnetic reconnection or internal shock) taking place below the transparency radius, the observed correlation can be reproduced.
 
%%%%%%%%%%%%%%%%%%%%%%%%%%%%%%%%%%%%%%%%%%%%%%%%%%%% 
 \subsubsection{The peak spectral energy-isotropic luminosity correlation $E_{peak}-L_{iso}$}
 %%%%%%%%%%%%%%%%%%%%%%%%%%%%%%%%%%%%%%%%%%%%%%%%%%%%%
The $E_{peak}$ and  $L_{iso}$  has been discovered \cite{109} with a sample of $16$ GRBs. A wider sample of 25 GRBs confirmed this
correlation \cite{56}.
 As  already said, the luminosity $L_{iso}$ is defined by combining the time-integrated spectrum of the burst with its peak flux (also $E_{peak}$ is derived using the time-integrated spectrum).  It has been demonstrated  that GRBs are characterized by a considerable spectral evolution  \cite{68}. If the peak luminosity is obtained only considering the spectrum integrated over a small time interval ($\sim 1 s$) centered around the peak of the burst light curve, we find a larger dispersion of the $E_{peak}-L_{iso}$ correlation (see \cite{108}). This suggests that, the time averaged quantities ({\it i.e.} the peak energy of the time-integrated spectrum and the "peak-averaged" luminosity) are better correlated than the "time-resolved" quantities.
 Combining the lag-luminosity and the variability- luminosity relations of nine GRBs to build their Hubble diagram it is possible to consider GRBs as powerful cosmological tools  \cite{S03,perillo}.

%%%%%%%%%%%%%%%%%%%%%%%%%%%%%%%%%%%%%%%%%%%%%%%%%%%%%%
 \subsubsection{The peak energy-collimation corrected energy correlation $E_{peak}-E_{\gamma}$}
 %%%%%%%%%%%%%%%%%%%%%%%%%%%%%%%%%%%%%%%%%%%%%%%%%%%%%%%%%%

By using a large sample of burst with spectroscopical measured redshift,
assuming a homogeneous circumburst medium, Ghirlanda et al. 
\cite{34} estimated the jet opening angle $\theta_{jet}$ and the
related $E_{\gamma}$ defined as $E_{iso} (1- \cos \theta_{jet} )$.
They discovered a very tight correlation between $E_{\gamma}$ respect
$E_{peak}$.  Such correlation links the GRB prompt emission energy,
corrected for the burst geometry, to its peak frequency. By adding a
large sample of GRB, some of which also discovered by SWIFT
\cite{swift} and, at least, by another InterPlanetary Network (IPN)
satellite \cite{IPN} in order to estimate the peak energy, this correlation has
been confirmed.  Recently, Nava et al., in \cite{35}, reconsidered and
updated the original sample of GRBs with firm estimate of their
redshift, spectral properties and jet break times.  One possible
explanation to see these events out of their jet opening angle in such
way that they appear off-beam-axis in both $E_{peak}$ and $E_{iso}$ has
been proposed (see also \cite{112,113}). Otherwise, alternative
possibility \cite{114,115} have been put forward.  Nava et al.
\cite{35} have calculated $E_{peak}-E_{\gamma}$ correlation by
assuming an external medium distributed with an $r^{-2}$ density
profile, finding an even tighter correlation with respect to that one
obtained assuming an homogenous circumburst density. The linearity of
the Ghirlanda $E_{peak}$ correlation implies that it is invariant
under transformation from the source rest frame to the fireball
comoving frame. A consequence of this property is that the number of
total photons emitted in different GRBs is constant and correspond to
$N_{\gamma}\sim 10^{57}$. This number is very close to the number
of baryons in one solar mass. This property could have an important
implication in the dynamics and radiative processes of GRBs.
 
 %%%%%%%%%%%%%%%%%%%%%%%%%%%%%%%%%%%%%%%%%%%%%%%%%%%%%%%%%%%%%%%
 \subsubsection{The isotropic luminosity-peak energy-high signal timescale correlation $L_{iso}-E_{peak}-T_{0.45}$ }
 %%%%%%%%%%%%%%%%%%%%%%%%%%%%%%%%%%%%%%%%%%%%%%%%%%%%%%%%%%%%%%%%%%%
 
 A correlation involving  three observables of the GRB prompt emission is discussed in \cite{fir06}:  the isotropic luminosity $L_{iso}$, the rest frame peak energy $E_{peak}$ and the rest frame "high signal" timescale $T_{0.45}$. This latter parameter has been previously used to characterize the GRB variability ({\it e.g.} \cite{89}) and represents the time crosses by the brightest $45\%$ of the total counts above the background. Through the analysis of $19$ GRBs, for which $L_{iso}$, $E_{peak}$ and $T_{0.45}$ could be derived, in \cite{fir06} it is found that $L_{iso}\propto E_{peak}^{1.62} \cdot T_{0.45}^{-0.49}$ with a very small scatter.
The $L_{iso}-E_{peak}-T_{0.45}$ correlation is based on prompt emission properties only and it  is sufficiently tight to standardize GRB energetics having some
interesting consequences: 
 \begin{enumerate}
 \item it represents a  powerful (redshift) indicator for GRBs without measured redshifts, which can be computed only from the prompt emission data (spectrum and light curve); 
 \item  it  can be considered a   cosmological tool  which is model independent (differently from the $E_{\gamma}-E_{peak}$correlation which relies on the standard GRB jet model \cite{fir06}); 
 \item it is "Lorentz invariant" for standard fireballs, {\it i.e.} when the jet opening angle is $\displaystyle{\theta_{jet} > \frac{1}{\Gamma}}\,.$
 \end{enumerate}
 These features could be extremely interesting in view of using GRBs as cosmological distance indicators.

%%%%%%%%%%%%%%%%%%%%%%%%%%%%%%%%%%%%%%%%%%%%%%%%%%%%%%%%%%%   
\subsubsection{The peak energy-isotropic energy-jet break time correlation $E_{peak}-E_{iso}-t_{break}$}
%%%%%%%%%%%%%%%%%%%%%%%%%%%%%%%%%%%%%%%%%%%%%%%%%%%%%%%%%%%%%

All the above correlations have been derived assuming that GRBs emit
isotropically. However, the hypothesis that GRB fireballs are
collimated was proposed for GRB $970508$ \cite{60} and subsequently
for GRB $990123$ as a possible explanation of their very large
isotropic energy \cite{61}. The collimated GRB model predicts the
appearance of an achromatic break in the afterglow light curve which,
after this break time, decreases with respect to the time
\cite{62,63}. Since the fireball photon emission depends on the
relativistic beaming angle, the observer receives the photons within a
cone with aperture $\theta_{\Gamma}\propto\frac{1}{\Gamma}$, where
$\Gamma$ is the bulk Lorentz factor of the material responsible for
the emission.  During the afterglow phase, the fireball is decelerated
by the circumburst medium and its bulk Lorentz factor decreases, {\it
  i.e.} the beaming angle $\theta_{\Gamma}$ increases with time. As
the beaming angle equals the jet opening angle,
$\theta_{\Gamma}\sim\frac{1}{\Gamma}\sim\theta_{jet}$, a critical time
is reached.  With  this assumptions, the jet opening angle
$\theta_{jet}$ can be estimated through this characteristic time
\cite{63}, {\it i.e.} the so called jet-break time $t_{break}$ related
to the afterglow light curve. Typical $t_{break}$ values ranges
from $0.5$ to $6$ days \cite{34, 58, 59}.  The jet opening angle
can be derived from $t_{break}$ in two different scenarios ({\it e.g.}
\cite{111}). Another empirical correlation links the jet break time $t_{break}$
with  $E_{iso}$ and peak energy $E_{peak}$.

 As discussed in \cite{35}, the model dependent $E_{peak}-E_{\gamma}$ correlations ({\it i.e.} derived under the assumption of a standard uniform jet model and either for a uniform or a wind circumburst medium) are consistent with this completely empirical  correlation. This result, therefore, reinforces the validity of the scenario within which they have been derived, {\it i.e.} a relativistic fireball with a uniform jet geometry which is decelerated by the external medium, with either a constant or an $r^{-2}$ profile density.
Similarly to what has been done with the isotropic quantities, we can explore if the collimation corrected $E_{peak}-E_{\gamma}$  correlation still holds when the luminosity, instead of the energy, is considered.

 %%%%%%%%%%%%%%%%%%%%%%%
 \subsection{Cosmological  analysis of GRB correlations}
 %%%%%%%%%%%%%%%%%%%%%

A preliminary step in the analysis of  correlations mentioned above
is the determination of the luminosity $L$ or the collimated energy
$E_{\gamma}$ entering as $y$ variable in the $y$\,-\,$x$ scaling laws
$\log y= a\log x+b $.  This is a linear relation which can be fitted
in order to determine the calibration parameter $a$ and $b$. Since,
there is still no theoretical model explaining any correlations in
term of GRB physics, one would expect that the wide range of GRB
properties makes the objects scatter around this (unknown) idealized
model. Consequently, the above linear relations will be affected by an
intrinsic scatter $\sigma_{int}$ which has to be determined together
with the calibration coefficients $(a, b)$.  As a first step, it is
necessary to determine the GRBs luminosity distance over a redshift
range where the linear Hubble law does not hold anymore. In such a way
the luminosity distance can be estimated as\,:

\begin{equation}
d_L(z) = \frac{c (1 + z)}{H_0} \int_{0}^{z}{\frac{dz'}{E(z')}}
\label{eq: dldef}
\end{equation}
where $H_0 = 100 h \ {\rm km/s/Mpc}$ is the present day Hubble constant
and $E(z) = H(z)/H_0$ is the dimensionless Hubble parameter which depends
on the adopted cosmological model thus leading to the well known {\it
circularity problem}.

Several strategies have been considered to take in to account this
problem \cite{perillo}. The most easy one is to assume a fiducial
cosmological model and determine its parameters by fitting, {\it
  e.g.}, the SNeIa Hubble diagram.  One adopts the $\Lambda$CDM model
as fiducial one thus setting\,:

\begin{equation}
E^2(z) = \Omega_M (1 + z)^3 + \Omega_{\Lambda}
\label{eq: ezlcdm}
\end{equation}
where $\Omega_{\Lambda} = 1 - \Omega_M$ because of the spatial flatness assumption. 

 It is nevertheless worth stressing that a different cosmological
 model would give different values for $d_L(z)$ and hence different
 values for $(L, E_{\gamma})$ thus affecting the estimate of the
 calibration parameters $(a, \sigma_{int})$. A first attempt to a
 model independent approach is to resort to cosmography
 \cite{Izzo08,W72,V04}, {\it i.e.}, to expand the scale factor $a(t)$ to the
 fifth order and then consider the luminosity distance as a function
 of the cosmographic parameters. Indeed, such a kinematic approach
 only relies on the validity of assumption of the Robertson\,-\,Walker
 metric, while no assumption on either the cosmological model or the
 theory of gravity is needed since the Friedmann equations are never
 used.  A further step to a completely model independent approach is
 the estimation of the GRBs luminosity distances by using the SNeIa as
 distance indicator based on the naive observations that a GRB at
 redshift $z$ must have the same distance modulus of SNeIa having the
 same redshift \cite{perillo}. As instance one could interpolate the
 SNeIa HD providing the value of distance modulus for a subsample of
 GRBs with redshift $z \le 1.4$ which can then be used to calibrate
 the correlations parameters \cite{K08,L08,WZ08,perillo}. Assuming
 that this calibration is redshift independent, one can then build up
 the HD at higher redshifts using the calibrated correlations for the
 remaining GRBs in the sample.

Once the calibration parameters for a certain  $y$\,-\,$x$ correlation have been obtained, it is then possible to estimate the distance modulus of a given GRB from the measured value of $x$. Indeed, for a given $y$, the luminosity distance is\,:

\begin{equation}
d_L^2(z) = \frac{y}{\kappa}\,,
 \end{equation}
 
 where $\kappa = 4 \pi P_{bolo}$, {\bf $\kappa = 4 \pi S_{bolo}
   F_{beam}/(1 + z)$} ($F_{beam}$is the beaming factor $(1-\cos\theta_{jet})$ and $\kappa = 4 \pi S_{bolo}/(1 + z)$ for $y =
 L$, $y = E_{\gamma}$ and $E_{iso}$, respectively.  Using the
 definition of distance modulus $ \mu(z) = 25 + 5 \log{d_L(z)}$ and
 estimating $y$ from $x$ through the $y$\,-\,$x$ correlation, we then
 get\,:

 \begin{eqnarray}
 \mu(z)  =  25 + \frac{5}{2} \log{\left(\frac{y}{\kappa}\right)} 
 =  25 + \frac{5}{2} (a \log {x} + b - \log{\kappa})
\label{eq: muval}
\end{eqnarray}
where $(a, b)$ are the best fit coefficients for the given $y$\,-\,$x$
correlation.  Eq.(\ref{eq: muval}) allows to compute the central value
of the distance modulus relying on the measured values of the GRB
observables, {\it i.e.}, the ones entering the quantity $\kappa$, and
the best fit coefficients $(a, b)$ of the used correlation. Moreover,
each correlation is affected by an intrinsic scatter which has to be
taken into account in the total error budget.

\begin{figure}
\centering
\includegraphics[width=10cm]{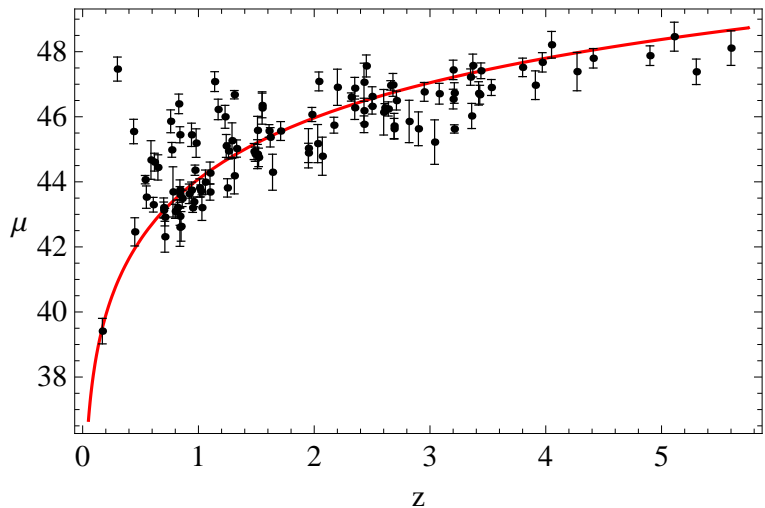}
\caption{GRB Hubble diagram  averaged over different 2D correlations derived by using calibration on the fiducial $\Lambda$CDM model \cite{perillo}.}\label{fig:1}
\end{figure}

In Fig.\,\ref{fig:1}, it is reported the
GRB Hubble diagram obtained by using different correlations and
different calibration methods (for  details see
\cite{perillo}).  The red solid line is the expected $\mu(z)$ curve
for the fiducial $\Lambda$CDM model.

From the picture, it is evident that the  HD derived from GRBs reasonably follow the $\Lambda$CDM curve although with a non negligible scatter.  The scatter becomes significantly larger in the range $0.4 \le z \le 1.4$ since the distance modulus $\mu(z)$ for a set of GRBs lies above the $\Lambda$CDM prediction. One could conjecture a failure of the theoretical model, but there is a set of points which are difficult  to adapt with any reasonable dark energy model.

%%%%%%%%%%%%%%%%%%%%%%%%%%%%%%%%%%%%%%%%%%%%%%%%%%%%
\section{Implications for particle astrophysics}
%%%%%%%%%%%%%%%%%%%%%%%%%%%%%%%%%%%%%%%%%%%%%%%%%%%%
As said above, a part electromagnetic emission, GRBs are sources of high-energy beams of particles and possible gravitational wave emitters. Below we will sketch these important implications for astroparticle physics and the possibility to complete the GRB light curve by studying radio emissions.

%%%%%%%%%%%%%%%%%%%%%%%%%%%
\subsection{High energy neutrinos}\label{subsec:fig}
%%%%%%%%%%%%%%%%%%%%%%%%%%%%%

According to  the fireball model, shock accelerated protons
interact with low energy radiation (photons in the range of KeV-MeV),
leading to the production of ultrahigh energy pions, kaons and
neutrons through the processes $p\gamma\rightarrow\pi^{\pm,0}X$ and $ p\gamma\rightarrow K^{\pm,0}X$, where $X$ can be a neutron. 
The ultrahigh
energy mesons and neutrons, in turn, decay into high energy
neutrinos/antineutrinos or photons and other secondary products.
Because of the high magnetic field inside GRBs, charged pions, kaons
and muons loose energy significantly before decaying into neutrinos.
On the other side, the relativistic neutrons remain unaffected and
decay into high energy antineutrinos, protons and electrons.

\begin{figure}[h!]
\centering{
\includegraphics[width=12.5cm]{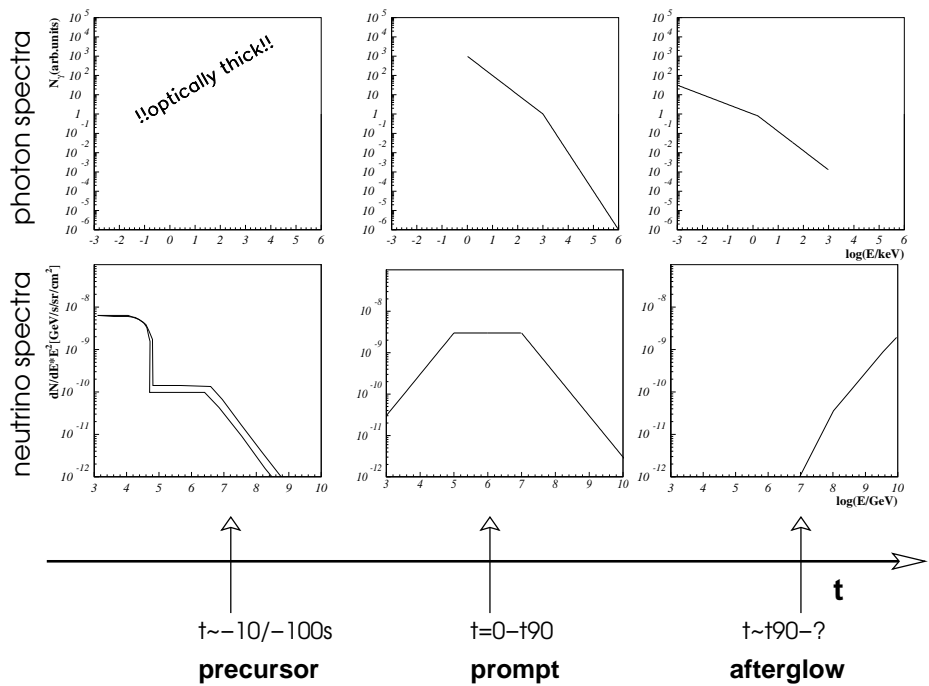}
\caption{Overview of different neutrino production scenarios during the three
  different phases of a GRB emission. The corresponding electromagnetic output is
  indicated schematically as well. The different flux models are described in
  the text \cite{Becker}.}
\label{grb_spectra:fig}}
\end{figure}

Three phases of non-thermal emission are expected: precursor phase preceding the
burst~\cite{precursor_nus}, the prompt emission~\cite{Waxman2,Waxman3} and
afterglow emission~\cite{afterglow_nus}. 
Fig.~\ref{grb_spectra:fig}  shows a schematic view of the modeled neutrino emissions from GRBs during the three phases (bottom row).
Also the corresponding electromagnetic outputs are shown (top row).

\begin{itemize}
\item A {\em precursor SNR (supernova remnant) model} of GRBs has
been developed in \cite{precursor_nus}. The basic idea is that a shock is formed 
 during the collision between the pre-GRB matter and the wind of the central pulsar or the
SNR shell.
In this state, the
burst is opaque to photon emission and shock surroundings represent a good
target for neutrino production by shock-accelerated protons interacting with
thermal $X$-rays in the sub-stellar jet cavity. The shocks happen at smaller
radii with respect to the prompt emission and at lower Lorentz boost factors $\Gamma$. The
neutrino signal could also be accompanied by a signal in the far infrared. The low
energy part of the neutrino spectrum is due to neutrino production in $p\,p$
interactions and follows the power law ${E_{\nu}}^{-2}$. 
 The neutrinos from the relativistic jet cavities are emitted as precursors ($\sim 10\div100$s prior) to the neutrinos emitted from the GRB fireball in case of electromagnetically observed  burst  \cite{Becker}. In case of electromagnetically undetectable burst it is not possible to detect neutrinos from individual sources since a diffuse neutrino signal is spread all around. 
\item The {\em prompt} {\em emission} from GRBs can be correlated to
  the observed flux of Ultra High Energy Cosmic Rays (UHECRs), since protons are accelerated  in
  the highly relativistic shocks by Fermi mechanism ~\cite{Vietri,Waxman2,Waxman3}. Such process
   implies the neutrino production through photon-hadronic interactions. 
   \item  {\it Afterglow neutrinos} are produced during the interaction between internal shocks from the
  original fireball and the interstellar medium \cite{afterglow_nus}. Neutrinos are produced in the external reverse shock due to the interaction of shock accelerated protons with synchrotron soft $X$-ray photons.  
\end{itemize}

In the GRB jet, a considerable number of neutrons ($n_n \simeq n_p$) is
expected, arising from a neutronized core similar to that in
supernovae in the case of long GRBs, and from neutron star material in
the case of short GRBs. In a long GRB,  the core collapse neutronization
leads to huge amount of thermal neutrinos ($\sim$ 10 MeV). Due to their low
energy, their cross section is too small for to be detected at
cosmological distances. However, in both long and short GRB outflows,
neutrons are present and initially coupled to protons by nuclear
elastic scattering. If the initial acceleration of the fireball is
very high, neutrons can eventually decouple from the fireball when the
comoving expansion time falls below the nuclear scattering
time. Protons, on the other hand, continue accelerating and expanding
with the fireball as they are coupled to electrons by Coulomb
scattering.  The flux and the spectrum of EeV neutrinos depends on the
density of the surrounding gas, while TeV-PeV neutrinos depend on the
fireball Lorentz factor. Hence, the detection of very high energy
neutrinos would provide crucial constraints on the fireball parameters
and on GRB environment.  Lower energy ($< $TeV) neutrinos originating
from sub-stellar shocks, on the other hand, may provide useful
information on the GRB progenitor.  

\noindent
Many experiment are searching neutrinos by GRBs.  The AMANDA search,
relied on spatial and temporal correlations with photon observations
of other instruments, such as BATSE, CGRO \cite{CGRO}  and other satellites of the
Third InterPlanetary Network (IPN) \cite{IPN}, leads to an increase in
sensitivity to a level where a single detected neutrino from a GRB can
be significant at the $5\sigma$ level \cite{AMANDA}. The neutrino flux
limit from AMANDA data as presented in~\cite{kyler_paper} is
consistent with the flux predicted in~\cite{Waxman3}:
\begin{equation}
\Phi^{DL}=6\cdot 10^{-9}\,GeV\,cm^{-2}\,s^{-1}\,sr^{-1}.
\end{equation}
The expected number of detected events from a given GRB is quite
low, however burst parameters can vary significantly from burst to
burst leading to a large variation in the expected number of detected
events. Thus the individual analysis of data from exceptionally bright
GRBs is highly interesting.  The AMANDA detector searched for neutrino
emission from more than $400$ GRBs, during the $1997\div 2003$ data
taking. The coincidence time was assumed to be the whole emission
duration over an excess of the background, while simulation-based data
quality cuts were applied to separate the predicted signal from the
observed background events. Zero neutrino events were observed in
coincidence with the bursts, resulting in a stringent upper limit on
the muon neutrino flux from GRBs
~\cite{icrc05_amanda,amanda_tev_proc,andreas,grbs_cascade07,5yrs}.
IceCube \cite{ice} performed searches for neutrinos from GRBs with two
analyses. In the first case, as for AMANDA, the selection of events
was made using IPN-$3$-detected bursts as triggers on the prompt
$\gamma$-ray emission of GRBs. However, since this kind of search
misses all GRBs which are not in the field of view of satellites or do
not emit gamma rays, a second analysis, seeking for unexpected
temporal clustering of events was performed. In this analysis, a time
window with a width fitting of the expected neutrino emission duration
(typically between $\,1s$ and $100\,s$) is imposed over the detected
events comparing the observed number of events with those expected for
the background.  In 2008 IceCube, in a 22-string configuration,
performed an analysis of individual GRB 080319B, the brightest GRB
ever recorded, finding no significant neutrino excess above the
background leading to the $90\%$ upper limit neutrino flux \cite{ice}. However
even for a $10$ times larger telescope the number of expected event is
only of the order of $1$.  Therefore, since the mean number of
expected neutrinos from individual GRBs is usually small, IceCube
performed a stacked GRB search, consisting in stacking several GRB
events coming from different directions. In this way, the chance for
discovery is noticeably enhanced, however it is no more possible to
associate the neutrino flux to a specific GRB.  An analysis with such an
approach was performed with IceCube data: the neutrino spectra were
calculated for $117$ bursts (mainly detected by SWIFT and FERMI \cite{fermi}) that
occurred during data taking. Even in this case the data analysis
showed no excess above the background.

The RICE experiment had investigated five bursts with respect to a
possible flux connected to the afterglow emission~\cite{rice_grbs07}, with
limits of a few orders of magnitude above the prediction, derived for each burst
individually \cite{afterglow_nus}. Stacking more
bursts, it can be possible to improve such limits. 

The SuperKamiokande Collaboration performed a search for neutrinos based on a time and direction correlation analysis with BATSE GRB solar neutrinos, atmospheric neutrinos, and upward-going muon data samples aimed at the search for an excess in the number of events correlated with GRBs above the expected background~\cite{Super}. Even in this case no statistically significant coincidences
were observed and only an upper limit on the fluence of neutrino-induced muons has been set.

An analysis aimed to  the detection of neutrino-induced showers in coincidence with  GRBs using the ANTARES detector, complementary to the track searches typically performed in neutrino telescope experiments, was performed \cite{antares}. This method has a lower sensitivity per burst with respect to the track search, however it can detect neutrinos of any flavor, which are invisible for track based analysis, and it is caable of looking also for coincidences with GRBs occurring in the northern hemisphere, as it does not depend on the direction of the observation. Ten GRBs triggered during the year $2008$ have been analyzed and  no a signal event were observed within the time-window. An upper limit on the normalization of this flux has been established \cite{antares}.

Recently, the GRB fireball neutrino flux calculation has been revised
taking into account the full spectral (energy) dependencies of the
proton and photon spectra, as well as the cooling of the secondaries,
flavor mixing, and additional multi-pion, kaon, and neutron production
channels for the neutrinos \cite{Winter}. A significant deviation has
been found in the normalization of the neutrino flux prediction of
about one order of magnitude, with a very different spectral shape
peaking at slightly higher energies. With this approach, neutrino flux
prediction is significantly below the current IceCube limit, which means
that the conventional GRB fireball phenomenology is not yet
challenged.

\begin{figure}
\centering{
\includegraphics[scale=0.65]{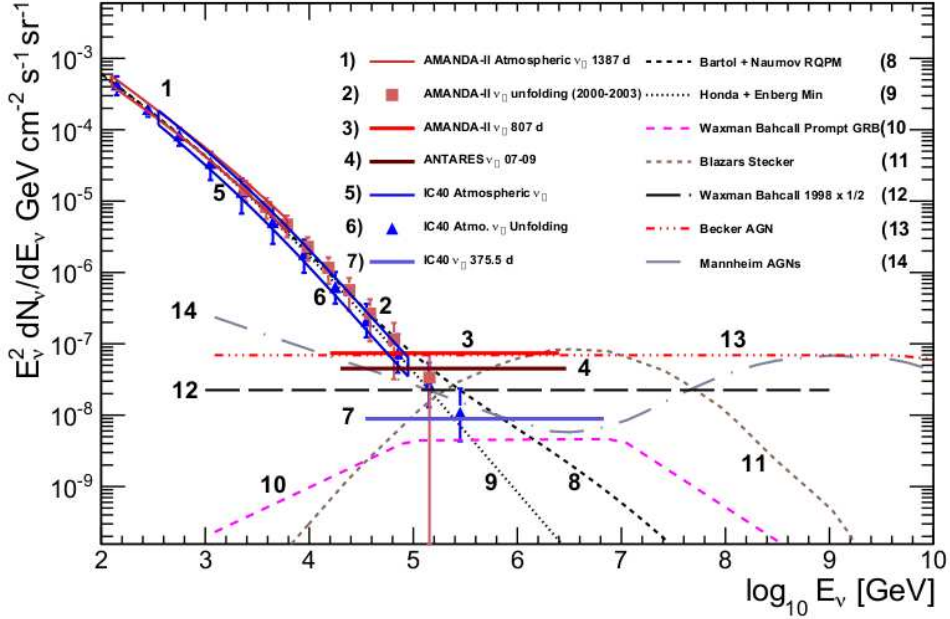}}
\caption{\label{singleflavorlimit}Upper limits on an astrophysical $\nu_{\mu}$ flux with an $E^{-2}$ spectrum are shown along with theoretical model predictions of diffuse astrophysical muon neutrinos from different sources.  The astrophysical $E^{-2} \ \nu_{\mu}$ upper limits  are from AMANDA-II, ANTARES, and the current work utilizing the IceCube 40-string configuration (IC40 $\nu_{\mu} \ 375.5$ d).  The atmospheric $\nu_\mu$ measurements are from AMANDA-II,  the IceCube 40-string (IC40) unfolding measurement and the current work (IC40 Atmospheric $\nu_{\mu}$) (\cite{DIFFUSE} and references therein).  }  
\end{figure}

\subsection{Gamma Ray Bursts and Gravitational waves}\label{subsec:prod}
%%%%%%%%%%%%%%%%%%%%%%%%%%%%%%%%%%%%%%%

The main GRB models  invoke a rapidly accreting black hole,
formed during a violent event such as the core collapse of a massive
compact star or the merger of two inspiraling binary neutron stars
associated to a strong emission of gravitational wave signals.
Whereas the $\gamma$-rays are thought to be produced at distances
$10^{13}$ cm from the central engine, these gravitational waves will
be produced in or near the central engine, and thus could provide our
most direct probe of it. For example, collapse or merger models lead
to different gravitational wave (GW)-burst energies, spectra, and polarizations
\cite{Meszaros:2006rc,meszaros03,razzaque,maggiore,tutukov}. Alternatively, GW production, owing to
toroidal instabilities in an accretion disk, will be relatively
long-lived and quasi-periodic, with an energy output of several orders of
magnitude higher than in the accretion mechanism was proposed in
\cite{Meszaros:2006rc}. In each case, the relative  arrival time of
GWB and GRB signals will depend on whether the GRB is generated by
internal shocks in the exploding fireball or external shocks when the
fireball is decelerated by the interstellar medium.  In addition, GRBs
occur at the rate of almost 1 observable event per day and are well
localized in time and, very often, on the sky, to  follow up GW
searches.  They are considered among the most promising GWs sources
for ground-based interferometers, of the current and future
generation, such as LIGO (Laser Interferometer Gravitational-Wave
Observatory) \cite{LIGO,LIGO1} and VIRGO \cite{VIRGO} and
their advanced versions.  
Although, with the current detector
sensitivities, the cosmological distances of most GRBs would lead to
individually undetectable GW signals, it is possible to accumulate
weak signals from several GRBs to detect a GW signature associated
to the GRB population.  Several Authors have described how the
signals from two independent GW-detectors can be analyzed to identify
the GW signals associated to GRBs and either bound or measure the
population average of the GW-flux on Earth from this potential
source. The key idea is to compare the mean correlated output of the
detectors during GRB events ("on" times) to the mean correlated output
when no GRB is detected ("off" times). Since gravitational waves from
GRBs would produce small correlations in the output of the two
detectors, a statistically significant difference in the mean
correlated output between on and off times would constitute an
indirect detection of GW-burst from GRBs. Alternately, the absence of a
statistically significant difference would allow one to set an upper
limit on the strength of any gravitational waves associated with GRBs.
Long-duration GRBs is thought to follow the star formation rate of the
Universe, and recent redshift measurements tend to support this model,
with the measured GRB redshift distribution peaking at $z\simeq
1$. Long-duration GRBs have also been associated exclusively with
late-type star-forming host galaxies. On the other hand, the recent
observations of $X$-ray and optical afterglows from a few
short-duration bursts seem to suggest that these GRBs are located at
lower redshifts relative to long-duration GRBs, and that short bursts
are found in a mixture of galaxy types, including elliptical galaxies,
which have older stellar populations. Although a large fraction of
GRBs are too distant for any associated GW signals to be detected by
LIGO, it is plausible that a small fraction occur at closer
distances. For example, a redshift of $z\,=\,0.0085$, {\it i.e.}   a distance of
$35$ Mpc, has been associated with long-duration burst/supernova GRB
980425/SN \cite{GRBSN}.

 It is reasonable to expect that a few GRBs with no
measured redshifts could be located relatively nearby as well.
Long duration GRBs can last over several tens of seconds, with the
time of peak flux appearing later than the time of first detection
of  GRB emission. This fact becomes important in the case of the
internal shock models for GRB emissions since the delay of $0.5$ sec
between the GRB and the GW signal that is expected to be
much smaller than the duration of the $\gamma$-ray light curve itself.
In case of a long GRB, due to the missing of an accurate prediction
of signal, analysis technique implemented by GW-detectors do
not rely on a detailed prediction on the waveform, but only impose
general bounds on signal duration and frequency.  

For short GRBs, the
redshift observations have led to fairly optimistic estimates
\cite{A34,A35} for an associated GW observation in an extended LIGO
science run.  In fact  observations support the hypothesis that
a large fraction of short GRBs can be produced by
NS-NS or NS-BH coalescence.  Since, GWs measurements of well-localized
inspiraling binaries can measure absolute source distances with high
accuracy, simultaneous observations of GWs (emitted by binary systems)
and short GRBs would allow us to directly and independently determine
both the luminosity distance and the redshift of  binary systems
(see \cite{Formisano} and references therein).  The combined
electromagnetic and gravitational view of these objects is likely to
teach us substantially more than what we learn from either data
channel alone. In \cite{Formisano}, the chirp masses associated to a
sample of short GRBs (under the hypothesis that they are emitted by
binary systems whose redshifts can be estimated considering them
comparable to the GRB redshift) is obtained. In such a way,
considering the coalescing time of the order of $T_{90}$-characteristic time\footnote{$T_{90}$ is defined as the time interval over which $90\%$ of the total background-subtracted counts are observed, with the interval starting when $5\%$ of the total counts have been observed.}   and the energy
lost (during the final phase of the coalescing process) equal to the
emission of short GRBs, it has been found that the chirp masses,
obtained by  simulations, are comparable to the theoretical chirp
masses associated to the coalescing binary systems.  Next generation
of interferometers (as LISA \cite{LISA}, or Advanced-VIRGO and LIGO)
could play a decisive role in order to detect GWs from these
systems. At advanced level, one expects to detect at least tens NS-NS
coalescing events per year, up to distances of order $2~$Gpc,
measuring the chirp masses with a precision better than $0.1\%$.

%%%%%%%%%%%%%%%%%%%%%%%%%%%%%%%%%%%%%%%%%%%%%%%%%%%%%%%%
\section{Gamma Ray Burst Radio emission}
%%%%%%%%%%%%%%%%%%%%%%%%%%%%%%%%%%%%%

In the fireball  model,  GRB afterglows are the result of a shock pushed into the ambient 
medium by an extremely relativistic outflow from
the GRB \cite{benz}.
It has been suggested that a possible signature of exploding fireball
is a burst of coherent radio emission \cite{Rees}.

The coherent radio emission is practically simultaneous
with the GRB except for the reduced propagation speed of
the radio waves by interstellar dispersion. A maximum delay of
a few seconds has been predicted in \cite{Palmer} for galactic
GRBs.
Another coherent radio emission produce a cloud of Compton electrons
propagating into the ambient magnetic field and radiating coherent
synchrotron emission for roughly one gyroperiod. Since the radiation
field of this emission process cannot exceed the ambient magnetic
field, it is undetectable at cosmic distances \cite{Katz}.  

In general, GRB afterglow observations are in good agreement with the
external shock scenario. Light curves at various wavelengths have been
obtained for  some bursts, and from these light curves, broadband
spectra have been constructed. For the broadband spectrum, synchrotron
radiation is assumed as the radiation mechanism. Dynamics of 
relativistic shock determines the spectrum evolution  with
time. The broadband synchrotron spectrum is determined by the peak
flux and three break frequencies, namely the synchrotron
self-absorption frequency $\nu_a$, the frequency $\nu_m$, that
corresponds to the minimal energy in the electron energy distribution,
and the cooling frequency $\nu_c$ that corresponds to electrons that
cool on the dynamical timescale.  

The break frequencies and the peak
flux can be described in terms of  energy of the blast-wave $E$, the
density of the surrounding medium $n$, the fractional energy densities
behind the relativistic shock in electrons and in the magnetic field,
$\epsilon_e$ and $\epsilon_B$ respectively, the slope $p$ of the
electron energy distribution, and the jet opening angle $\theta
_{jet}$.  Modeling radio emission provides the possibility to
investigate the immediate surroundings of the burst and the initial
Lorentz factor of blast-wave. The fact that the timescale for detecting
emission from the reverse shock is  small at optical and much longer
at radio wavelengths makes radio observations on the first day after
the burst important, although the emission from the forward shock is
maybe not detectable at those early times.  

The uniqueness of radio
afterglow observations is best illustrated by the phenomenon of radio
scintillation. Scintillation due to the local interstellar medium
modulates the radio flux of GRB afterglows and permits indirect
measurement of the angular size of the fireball.  Focusing and
defocusing of the wave front by large-scale inhomogeneities in the
ionized interstellar medium results in refractive scintillation \cite{frail}.  
This scintillation is broadband and has been
seen in many sources, whereas only the most compact sources, {\it
  e.g.} GRB afterglows, show diffractive scintillation. Diffractive
scintillation is caused by interference between rays diffracted by
small-scale irregularities in the ionized interstellar medium. The
resulting interference is narrow-band and highly variable.
Diffractive scintillation occurs only when the source size is smaller
than a characteristic size, the so-called diffractive angle. It turns
out that at an average redshift of $z\,\sim\, 1$, the size of the
blast-wave is smaller than the diffractive angle in the first few days,
but during its evolution the blast-wave becomes larger than this angle.
Thus, as the blast-wave expands, the diffractive scintillation is
quenched. By studying the scintillation behavior, one can get an
independent measurement of the angular size of the blast-wave. The
radio light curves are phenomenological different from optical and
$X$-ray light curves in the sense that the flux at radio wavelengths
first increases at a timescale of weeks to months before it starts to
decline.  Since $X$-ray afterglows are very weak, and optical
afterglow observations are contaminated by the presence of a host
galaxy or a possible supernova, radio afterglows provide essential
informations in late-time, following  GRBs in their evolution into
the non-relativistic phase. These late-time observations can give a
determination of the blast-wave energy independent of the initial jet
collimation.  Moreover they allow to observe phenomena which would
otherwise escape attention, such as the occurrence of a radio flare.

\begin{figure}
\centering
\includegraphics[height=6.85cm]{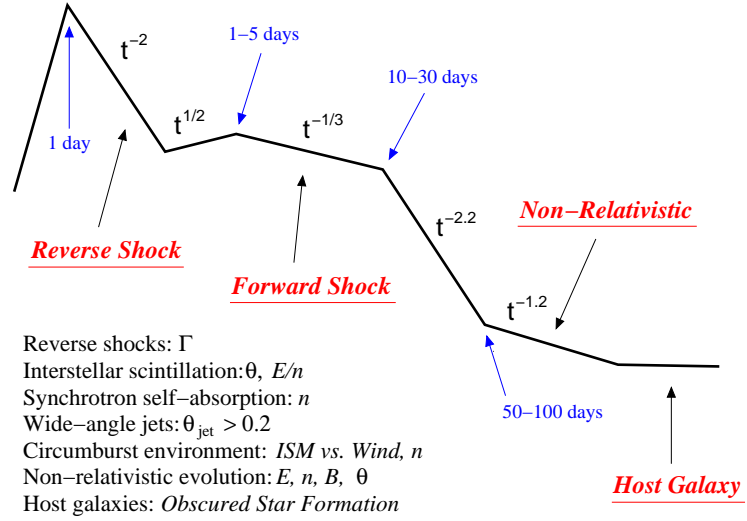}
\caption{A schematic radio afterglow light curve. Timescales and
  scalings for the temporal evolution are indicated.  The list
  summarizes aspects of the flux evolution which are unique to the
  radio bands (Lorentz factor, $\Gamma$; source size, $\theta$;
  energy, $E$; density, $n$; jet opening angle, $\theta_{\rm jet}$;
  density profile; magnetic field strength, $B$; and obscured star
  formation rate)\cite{frail}.
\label{fig:cartoon}}
\end{figure}

The evolution a GRB and its radio afterglow is modelled 
 schematically in Fig. \ref{fig:cartoon}. The observations cover four orders of magnitude
in time ($0.1\div1000$ days) and three orders of magnitude in frequency ($0.8\div660$ GHz), so that radio light curves exhibit a
rich phenomenology.

On a timescale of days to weeks after the burst, the subsequent evolution
of the radio afterglow (Fig. \ref{fig:cartoon}) can be described by a slow rise to maximum,
followed by a power-law decay. The radio peak is often accompanied by a
sharp break in the optical (or $X$-ray) light curves \cite{fir06, LZ}. 

The most commonly
accepted (but not universal) explanation for these achromatic breaks is that
GRB outflows are collimated. The change in spectral slope, 
where the flux density is $F\propto t^{\alpha}\nu^\beta$,
 occurs when the Lorentz factor $\Gamma$ of the shock drops below the inverse opening jet angle $\theta_{jet}$.

Since the radio emission initially lies below the
synchrotron peak frequency $\nu_m$, the jet break signature is distinctly different
than that at optical and $X$-ray wavelengths. 

Then jet break is expected to give rise to a shallow decay
$t^{-\frac{1}{3}}$ .  Another recognizable radio signature of a
jet-like geometry is the {\it peak flux cascade}, in which successively
smaller frequencies reach lower peak fluxes ({\it i.e.}
$F\nu_m\propto\sqrt{\nu_ m}$ ).

At sufficiently late times, when the rest mass energy swept up by the
expanding shock becomes comparable to the initial kinetic energy of
the ejecta ($\sim100$ days), the expanding shock may slow to
non-relativistic speeds \cite{piran04}. A change in the temporal slope
is expected (Fig. \ref{fig:cartoon}) for a constant density medium,
independent of geometry.  Finally, the radio light curves at late
times may flatten due to the presence of an underlying host
galaxy. Most GRBs studied so far have optical/NIR hosts but only about
$20\%$ have been seen at centimeter and submillimeter wavelengths
\cite{B48,B49}. This is an emerging area with a great potential of
study but requires a sensitivity that only a few radio telescopes
have, as Very Large Array (VLA) \cite{VLA}, Low Frequency Array
(LOFAR) \cite{LOFAR}, and the forthcoming Square Kilometer Array (SKA) \cite{SKA}.
Such instruments allow to investigate the GRB radio afterglows at high
redshifts including also the possibility of more energetic versions of
these objects that may be associated with Population III stars.  This
can be important for broadband afterglow fits aimed at determining
total energies and physical parameters of bursts, out to very high
redshifts.

%%%%%%%%%%%%%%%%%%%%%%%

\section{Conclusions}

GRBs are flashes of $\gamma$-rays associated with extremely energetic explosions that have been observed in distant galaxies. 
As discussed, they can be
roughly separated into two classes \cite{Weeeks}, long GRBs (with
T$_{90}$ $\gtrsim2$s), associated to gravitational collapse of
very massive stars, and short GRBs (with T$_{90}$ $\lesssim2$s),
associated to mergers of compact objects. 
GRBs have recently attracted a lot of attention as promising standardizable objects
candidates to describe the Hubble diagram up to very high $z$, deep into
the matter dominated era thus complementing SNeIa which are, on the
contrary, excellent probes for the dark energy epoch. However, still
much work is needed in order to be sure that GRBs can indeed hold this
promise.

Searching for a relation similar to that used to standardize SNeIa has
lead to different empirically motivated scaling relations for GRBs.
Anyway there are still open issues related to the use of GRBs in
cosmology:
\begin{itemize}
 \item the low number of
events: the samples of GRBs which can be used to constrain  cosmological parameters through
the discussed correlations are not so rich;
  \item  the absence
of GRBs at  low redshift does not allow to calibrate
the correlations and requires to adopt  methods to fit the cosmological parameters in order to avoid
the circularity problem.
\end{itemize}
  Moreover, the lack of  theoretical interpretation for the physics
 of  these correlations represents a still open issue.
  The increase of the number of bursts which can be used to measure the cosmological
parameters, and the possible calibration of the correlations would greatly improve the constraints
that are presently obtained with few events and  non-calibrated correlations.
In order to use GRBs as a cosmological tools, through the above correlations, three
fundamental parameters, {\it i.e.} $E_{peak}$, $E_\gamma$ and $\theta_{jet}$, should be accurately measured. On the other hand the $L_{iso}-E_{peak}-T_{0.45}$
does not require the knowledge of the afterglow emission because it completely relies on the
prompt emission observables. 
 The need to know the cosmological model to infer the luminosity
distance for each GRB contrasts with the desire to constrain that same
cosmological model (circularity problem).  In the attempt to overcome this problem, one can take into account 
  scaling relations and  derive the Hubble diagram by
 different methods in order to estimate the luminosity distance \cite{perillo}.
  
Moreover, GRBs are powerful sources of high-energy neutrinos emitted in different phases according to the fireball model .
A mechanism leading to higher (GeV) energy neutrinos in GRB is due to inelastic nuclear
collisions.
Proton-neutron inelastic collisions are expected, at much lower radii than radii
where shocks occur, due to the decoupling of neutrons and protons in the fireball or jet phases.
If the fireball has a substantial neutron/proton ratio, as expected in most GRB
progenitors, the collisions become inelastic and their rate peaks  where the nuclear
scattering time becomes comparable to the expansion time.
Inelastic neutron/proton collisions then lead
to charged pions, GeV muon and electron neutrinos.  
The typical GRBs neutrino energies 
range from multi-GeV to EeV, and can yield interesting physical information about fundamental interactions, about (ultra-high energy) cosmic rays, and about the nature of GRBs and their environment.
The GRBs neutrino signals may be  detected in the coming years by current and forthcoming experiments such as Ice-Cube, RICE, and KM3NeT \cite{KM3NeT}.
While the $\pi$ interactions leading to $>100 $TeV
energy neutrinos provide a direct probe of the internal shock acceleration process, as
well as of the MeV photon density associated with them, the $>10$ PeV neutrinos
would probe the reverse external shocks, as well as the photon energies and energy
density there.
 In the very recent years several neutrino telescopes are performing a systematic search for neutrinos emission from GRBs with different analysis methods. Up to now, no signal in excess over the background rate has been observed.
 
The leading models for the ultimate energy source of GRBs are stellar collapse or
compact stellar mergers, and these are expected to be sources of 
GWs. If some fraction of GRBs are produced by
double neutron star or neutron star-black hole mergers, the gravitational wave
chirp signal of the in-spiral phase should be detectable by the advanced LIGO-VIRGO, associated with the GRB electromagnetic signal.
Although the waveforms of the
gravitational waves produced in the break-up, merger and/or bar instability phase
of collapsars are not known, a cross-correlation technique can be used making use of
two co-aligned detectors.

The understanding  of GRB physics is  today rapidly advancing 
since the discovery of long-lived "afterglow" emission is giving a great insight into the problem. Radio
afterglow studies have become an integral part of this field, providing
complementary and sometimes unique diagnostics on GRB explosions, their
progenitors, and their environments. The reason for this is that the radio part
of the spectrum is phenomenologically rich, but also difficult to investigate because only $20\%$ of GRBs observed so far have been seen at radio-wavelength.  A GRB
radio-survey requires a very high sensitivity that only few radio telescopes can reach.  The forthcoming Square Kilometers Array  (SKA) could be of extreme interest in this effort.

\section*{Acknowledgements}
The Authors  acknowledge L. Amati,  G. Barbarino,  V.F. Cardone,  M.G. Dainotti,  T. Di Girolamo and M. Perillo for useful discussions and comments on  topics related with this paper.

%%%%%%%%%%%%%%%%%%%%%%%%%%%%%%%%%%%%%%%%%%%%%%%%%%
%%%%%%%%%%%%%%%%%%%%%%%%%%%%%%%%%%%%%%%%%%%%%%%%%%%%%%%%%%
%%%%%%%%%%%%%%%%%%%%%%%%%%%%%%%%%%%%%%%%%%%%%%%%

\end{document}